\documentstyle[seceq,epsf]{ptptex}

\newcommand{\I}{{\rm I}}
\newcommand{\II}{{\rm II}}
\newcommand{\tI}{\widetilde{\rm I}}
\newcommand{\tII}{\widetilde{\rm II}}
\newcommand{\bt}{\bar{t}}
\newcommand{\bx}{\bar{x}}

\newcommand{\bz}{\bar{z}}

\newcommand{\we}{{w_{\perp}}}

\newcommand{\X}{\bar{X}}

\newcommand{\rh}{\hat{\rho}}

\title{Classical Radiation Formula in the Rindler Frame}

\author{Toru {\sc Hirayama}\footnote{E-mail: hira@cc.kyoto-su.ac.jp}}

\inst{Department of Physics, Kyoto Sangyo University, 
Kyoto 603-8555, Japan}

\abst{
In a preceding paper 
$[$T.~Hirayama, Prog.~Theor.~Phys.~{\bf 106} (2001), 71$]$, 
the power of the classical radiation emitted by a moving charge was 
evaluated in the Rindler frame. 
In this paper, we give a simpler derivation of this radiation formula, 
including an estimation of the directional dependence of the radiation. 
We find that the splitting of the energy-momentum 
tensor into a bound part $\tI$ and an emitted part $\tII$ is 
consistent with the three conditions introduced in the preceding paper, 
also for each direction within the future light cone.
}

\begin{document}

\maketitle

\section{\label{introo}Introduction}

In a preceding paper,\cite{Hi01} we calculated the power of 
the classical elecromagnetic radiation emanated from a moving 
point charge in the Rindler frame, which is a linearly and uniformly 
accelerated frame in Minkowski spacetime and also interpreted as 
a rest frame with a static homogeneous gravitational field.\cite{Ri66} 
The power was evaluated in terms of the energy defined by the 
Killing vector field that generates the time of that frame. We 
found that the power is proportional to the square of the 
acceleration $\alpha^\mu$ of the charge relative to 
the Rindler frame $[$see Ref.~\citen{Hi01} 
or \S\ref{energyy} of this paper for the definition of $\alpha^\mu]$.
This result was interpreted as providing a picture in which the 
concept of radiation depends on the reference frames or motion of 
observers,\cite{PV99} and we discussed it in relation to the old 
paradox concerning the classical radiation emitted from a uniformly 
accelerated charge.\cite{FR60,Bo80}

In the conventional treatments, radiation is identified with 
reference to the region sufficiently distant from the charge. 
(In the case of the Rindler spacetime, this region 
would correspond to the future horizon.) Actually, the well-known
Larmor formula and the above mentioned radiation formula in the 
Rindler frame are obtained by integrating the energy in that 
region.
However, Rohrlich and Teitelboim proposed another 
approach,\cite{Ro61,Te70,TV80} in which the radiation can be 
identified at an arbitrary distance from the charge, without 
reference to the asymptotic zone. This provides a picture of 
radiation as an emission of something by a charge, which 
begins to exist immediately after emission. This approach 
consists of splitting the energy-momentum tensor of the
retarded field into an emitted part $\II$ and a bound 
part $\I$, which can be done in a natural way using the usual 
splitting of the retarded field into the acceleration part 
and the velocity part.\cite{Te70,TV80}

Similarly, we found in Ref.~\citen{Hi01} that a natural 
splitting of the energy-momentum tensor into an emitted part 
$\tII$ and a bound part $\tI$ exists in the Rindler frame. 
This splitting was realized by splitting the retarded field 
into a part linear in $\alpha^\mu$ and a part independent of 
$\alpha^\mu$. The emitted nature and bound nature of these parts 
were confirmed in the sense that they satisfy the 
following three conditions:\cite{Hi01}
\\

\begin{enumerate}

\item[1.]{A charge fixed in the Rindler frame does not radiate}.

\item[2.]{The emitted part propagates along the future light 
cone with an apex on the world line of the charge, 
in the sense that the energy of this part does not damp 
along the light cone}. 

\item[3.]{The bound part does not contribute to the energy in 
the region $\eta\rightarrow\infty$, where $\eta$ is the time 
coordinate in the Rindler frame and the limit is taken along 
the future light cone with an apex on the world line 
of the charge}.
\\

\end{enumerate}

The main purpose of this paper is to present a simpler 
derivation of the radiation formula in the Rindler frame by 
using retarded coordinates,\cite{TV80,Ta76} which are 
often used to simplify the calculation of retarded 
quantities around a charge. Simplification also results from the
relation expressed in Eq.~(\ref{simple}), in which we can 
characterize the distance between the charge and the future 
horizon in terms of the null vector $w^\mu$, the vector which played 
a key role in Ref.~\citen{Hi01}. In this calculation, the angular 
dependence of the radiation is also represented obviously in the 
integrand of the equation.

In Ref.~\citen{Hi01}, the consistency of splitting the 
tensor into $\tI$ and $\tII$ with three conditions was confirmed 
only after integration over the angular dependence. In this 
paper, we confirm this consistency for each solid angle element. 
This procedure is again 
simplified with the aid of Eq.~(\ref{simple}). In these 
calculations, we use the solid angle defined in the inertial 
frame. This is reasonable because the limit in which the radiation  
is identified should be taken along the light ray (null geodesic) 
that emanates from the charge, and the angular coordinates in 
the inertial frame are constant along this light ray. It is also 
pointed out that the radiative energy per solid angle is 
nonnegative, as would be expected in natural identification of 
radiation. 

Throughout this paper, we use Gaussian units and the metric with 
signature $(-,+,+,+)$.

\section{\label{energyy}Simple derivation of the radiation 
formula}

In this section, we calculate the power of the radiation in the 
Rindler frame by using retarded coordinates, which are 
defined as follows.\cite{TV80,Ta76} For a point $p$ (with 
coordinates $x^\mu$), the retarded point $\bar{p}$ 
(with coordinates $\bx^\mu$) is defined as a point on the  
worldline of a point charge that satisfies the 
condition\footnote{We use $r^\mu$ 
and $\rho$ throughout this paper, insted of $R^\mu$ 
and $R$ used in Ref.~\citen{Hi01}}
\begin{eqnarray}
r^\mu&=&x^\mu-\bx^\mu,\\
r^\mu r_\mu&=&0,\ \ \ \ x^t > \bx^t,
\end{eqnarray}
expressed in an inertial frame $(t,x,y,z)$. We define the 
retarded time $\tau$ of a point $p$ as the proper time of a charge 
at the retarded point $\bar{p}$. The retarded distance  $\rho$ 
is defined as $\rho=-v\cdot r$, where $v$ denotes the 4-velocity 
of the charge at the point $\bar{p}$. Then we can specify an 
arbitrary point in the spacetime by the coordinates $\tau$ and $\rho$ 
and the angular coordinates $\theta$ and $\varphi$ defined in an 
inertial frame.

We now introduce a ray vector $k^\mu=r^\mu/\rho$, which has the 
properties $k\cdot k=0$ and $k\cdot v=-1$. This vector does not 
depend on $\rho$, and it is expressed as
$k^\mu=k^\mu(\tau,\theta,\varphi)$. 
The volume element $d\Sigma_\mu$ of the 3-dimensional plane 
$\Sigma$ orthogonal to the vector $N^\mu$ is calculated in 
Appendix \ref{v_element}, and the result is
\begin{eqnarray}
d\Sigma^\mu&=&-\frac{N^\mu}{N\cdot k}\rho^2d\tau d\Omega_0,
\label{volume}
\end{eqnarray}
where $d\Omega_0$ is the solid angle element of the inertial 
frame with time axis $v^\mu$ $[$see Eq.~(\ref{solid})$]$.

For the inertial coordinates $(t,x,y,z)$, the Rindler coordinates 
$(\eta,x,y,\xi)$ are given by $t=\xi\sinh\eta$ and 
$z=\xi\cosh\eta$, with the metric 
$ds^2=-\xi^2 d\eta^2+dx^2+dy^2+d\xi^2$.
We can define the energy in the Rindler frame with respect to 
the time-like Killing vector field $X^\mu$ which is proportional 
to the field\footnote{We here note that the Killing field 
$X=\nu\partial/\partial\eta$
represents a time coordinate in the Rindler frame normalized by 
the proper time of a Rindler observer fixed at
$\xi=\nu^{-1}$.}
\begin{eqnarray}
\frac{\partial}{\partial\eta}&=&
z\frac{\partial}{\partial t}+t\frac{\partial}{\partial z}.
\end{eqnarray}
The Rindler energy over the region $\Sigma$ is given by 
\begin{eqnarray}
E_{[X,\Sigma]}&=&-\int_\Sigma d\Sigma_\mu X_\nu T^{\mu\nu},
\label{ESigma}
\end{eqnarray}
where  
$T^{\mu\nu}$ is the energy-momentum tensor of the electromagnetic 
field. For free fields, conservation of this quantity 
is confirmed by the Killing equation 
$\nabla_\mu X_\nu+\nabla_\nu X_\mu=0$
and Gauss's theorem $[$see \S 2.1 of Ref.~\citen{Hi01}$]$.

\begin{figure}[t]
\epsfxsize=9cm
\centerline{\epsfbox{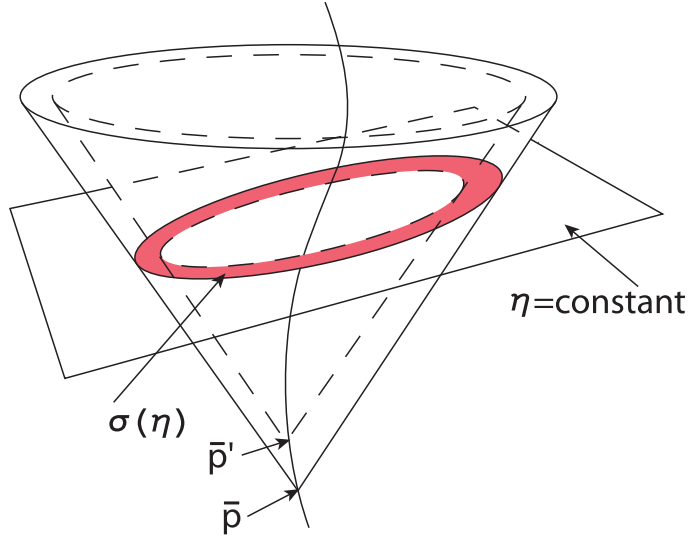}} 
\caption{\label{nullfig}The radiation in the Rindler frame is 
evaluated with respect to the energy in the region 
\protect$\lim_{\eta\rightarrow\infty}\sigma(\eta)$. 
To identify the emitted and bound parts of the Maxwell tensor 
properly, we introduce three conditions concerning the behavior 
of these parts along the future light cone.}
\end{figure}

We consider two closely positioned points $\bar{p}$ and $\bar{p}'$ 
on the worldline of the charge (see Fig.~\ref{nullfig}). Let 
$\Delta L$ be a region that is bounded on two future light 
cones with apexes $\bar{p}$ and $\bar{p}'$. Let $\sigma(\eta)$ 
be the intersection of  $\Delta L$ with a surface of 
simultaneity $\eta=[{\rm constant}]$. In the following, 
we determine the radiation emitted by a moving charge 
in the Rindler frame by evaluating the Rindler energy of the 
retarded field in the region 
$\sigma(\eta\rightarrow\infty)$.

We here note the relation
\begin{eqnarray}
X\cdot k=\X\cdot k, 
\label{Killing}
\end{eqnarray}
which is satisfied on the future light cone with the apex 
$\bar{p}$, where $\X^\mu$ is the value of $X^\mu$ evaluated at 
$\bar{p}$. We can prove this relation for general Killing fields 
in Minkowski spacetime as follows. We can write 
$k^\mu \nabla_\mu k^\nu=0$, because $k^\mu$ is independent of $\rho$. 
This equation and the 
Killing equation $\nabla_\mu X_\nu=-\nabla_\nu X_\mu$ lead to 
$k^\mu\nabla_\mu(X\cdot k)=0$. Thus $X\cdot k$ is constant 
along the ray vector that emanates from $\bar{p}$. 

The volume element for $\sigma(\eta)$ is obtained from  
Eq.~(\ref{volume}) by using $X^\mu$ for $N^\mu$. 
Therefore,  from Eqs.~(\ref{ESigma}) and (\ref{Killing}), the Rindler 
energy over the region $\sigma(\eta)$ can be expressed as
\begin{eqnarray}
E_{[X,\sigma(\eta),{\rm ret}]}&=&
d\tau\int_{\sigma(\eta)}d\Omega_0
\frac{\rho^2}{\X\cdot k}X_\mu X_\nu {T_{\rm ret}}^{\mu\nu},
\label{sigmaeta}
\end{eqnarray}
where ${T_{\rm ret}}^{\mu\nu}$ is the energy-momentum tensor of 
the retarded field generated by the charge. 
Except in the direction of $z$, or the $\xi$ axis, the region 
$\sigma(\eta\rightarrow\infty)$ arrives at the future horizon 
$H^+$. Therefore, we now derive two formulae giving the values of 
$\rho$ and $X^\mu$ at $H^+$, in order to evaluate the value 
of the integrand in Eq.~(\ref{sigmaeta}) in that region.

Let us consider a null vector 
$w^\mu=(g\cdot g)^{1/2}u^\mu+g^\mu$, which plays a key role 
in defining the bound and emitted parts of the Maxwell tensor 
in Ref.~\citen{Hi01}. Here 
$u^\mu=|X\cdot X|^{-1/2}X^\mu|_{\bar{p}}$ 
and $g^\mu=|X\cdot X|^{-1}X^\nu\nabla_\nu X^\mu|_{\bar{p}}$ 
represent the 4-velocity and the 4-acceleration of the Rindler 
observer at the point $\bar{p}$, respectively. We here note that 
throughout this paper, $w^\mu$ is considered at the point $\bar{p}$. 
In the inertial frame $(t,x,y,z)$, $w^\mu$ is written 
\begin{eqnarray}
(w^t,w^x,w^y,w^z)&=&
\frac{1}{\bz-\bt}(1,0,0,1).
\end{eqnarray}
The contraction of this quantity with the vector $r^\mu|_{H^+}$ 
gives
\begin{eqnarray}
r^\mu|_{_{H^+}}w_\mu&=&-1,
\label{simple}
\end{eqnarray}
which is obtained by substituting $z=t$, which holds for the 
future horizon $H^+$. To avoid confusion caused by our 
unfamiliar notation, we here note again that $w^\mu$ is 
evaluated at $\bar{p}$, not at $H^+$.

For an inertial frame with time axis $t^\mu$, where $t^\mu$ is
a future-directed vector and $t^\mu t_\mu=-1$, we introduce the 
retarded distance $\rh=-r^\mu t_\mu$ and the ray vector 
$l^\mu=r^\mu/\rh$. Equation~(\ref{simple}) divided by the 
retarded distance $\rh|_{H^+}$ gives
\begin{eqnarray}
\left.
\frac{1}{\ \rh\ }
\right|_{H^+}&=&-w^\mu l_\mu.
\label{formula1}
\end{eqnarray}
On the other hand, if a vector $w'^\mu$ satisfies   
$\rh^{-1}|_{H^+}=-w'^\mu l_\mu$, then $w'^\mu=w^\mu$, because we 
can choose four independent ray vectors $l^\mu$ that emanate 
from $\bar{p}$, and the contractions of these four vectors with 
$w'^\mu$ uniquely determine all components of the vector 
$w'^\mu$. From this equation, we find that the distance 
between the observer and the event horizon $H^+$ is 
charactarized by a unique vector $w^\mu$, and that this 
expression is the same for any inertial frame, i.e., 
independent of the choice of $t^\mu$.

For the inertial frame with the time axis 
$v^\mu$, Eq.~(\ref{formula1}) reduces to  
\begin{eqnarray}
\left.
\frac{1}{\ \rho\ }
\right|_{H^+}&=&-w^\mu k_\mu.
\label{horizon1}
\end{eqnarray}
Using this equation and noting that $X^\mu|_{H^+}\propto w^\mu$, we 
find that 
$X^\mu/\rho|_{H^+}=-(w\cdot k)X^\mu|_{H^+}
=-(X\cdot k)|_{H^+}w^\mu$.
From Eq.~(\ref{Killing}), we obtain the formula for evaluating 
$X^\mu$ at $H^+$,
\begin{eqnarray}
\left.
\frac{X^\mu}{\rho}
\right|_{H^+}
&=&
-(\X\cdot k)w^\mu.
\label{horizon2}
\end{eqnarray}

Now we can evaluate the radiative power in the region $H^+$ by 
using Eqs.~(\ref{horizon1}) and (\ref{horizon2}) in
Eq.~(\ref{sigmaeta}) in the limit $\eta\rightarrow\infty$, with  
the explicit form of ${T_{\rm ret}}^{\mu\nu}$ given as
$[$see Eq. (A$\cdot$17) in Ref.~\citen{Hi01} or Eq. (2.7) in 
Ref.~\citen{Te70}, and also Eq.~(\ref{Tei_splitting}) in this paper.$]$
\begin{eqnarray}
\hspace{-1cm}{T_{\rm ret}}^{\mu\nu}&=&
{T_{\I,\I}}^{\mu\nu}+{T_{\I,\II}}^{\mu\nu}
+{T_{\II,\II}}^{\mu\nu},
\label{separate}\\
\hspace{-1cm}\frac{4\pi}{e^2}{T_{\I,\I}}^{\mu\nu}
&=&
\frac{1}{\rho^4}
\left[
k^\mu k^\nu-(v^\mu k^\nu+k^\mu v^\nu)-\frac{1}{2}\eta^{\mu\nu}
\right],
\label{IandI}\\
\hspace{-1cm}\frac{4\pi}{e^2}{T_{\I,\II}}^{\mu\nu}&=&
\frac{1}{\rho^3}
[2(a\cdot k)k^\mu k^\nu-(a\cdot k)(v^\mu k^\nu+k^\mu v^\nu)
-(a^\mu k^\nu+k^\mu a^\nu)],
\label{IandII}\\
\hspace{-1cm}\frac{4\pi}{e^2}{T_{\II,\II}}^{\mu\nu}&=&
\frac{1}{\rho^2}
[(a\cdot k)^2-a\cdot a]
k^\mu k^\nu,
\label{IIandII}
\end{eqnarray}
where $v^\mu$ and $a^\mu$ are the 4-vector and the 4-acceleration 
of the charge evaluated at the retarded point $\bar{p}$.
First, let us consider the contribution from 
${T_{\I,\I}}^{\mu\nu}$. From Eqs.~(\ref{horizon2}) and 
(\ref{IandI}), we find
\begin{eqnarray}
\left.
\frac{\rho^2}{\X\cdot k}
\frac{4\pi}{e^2}{T_{\I,\I}}^{\mu\nu}X_\mu X_\nu
\right|_{H^+}
&=&
\left.
\rho^4(\X\cdot k)
\frac{4\pi}{e^2}{T_{\I,\I}}^{\mu\nu}w_\mu w_\nu
\right|_{H^+}
\nonumber\\
&=&
(\X\cdot k) w\cdot k(w\cdot k-2v\cdot w)\nonumber\\
&=&(\X\cdot k)
\left[
(w_\perp\cdot k)^2-(w_\perp\cdot w_\perp)
\right].
\label{hI_I}
\end{eqnarray}
Here ${w_\perp}^\mu=h^\mu_{\ \nu}w^\nu$, where 
$h^\mu_{\ \nu}=\delta^\mu_{\ \nu}+v^\mu v_\nu$
is the projector onto the plane orthogonal to $v^\mu$.
We find, similarly, for ${T_{\I,\II}}^{\mu\nu}$,
\begin{eqnarray}
\left.
\frac{\rho^2}{\X\cdot k}
\frac{4\pi}{e^2}{T_{\I,\II}}^{\mu\nu}X_\mu X_\nu
\right|_{H^+}
&=&
2(\X\cdot k)
\left[
a\cdot w_\perp-(a\cdot k)w_\perp\cdot k
\right]
\label{hI_II}
\end{eqnarray}
To determine the contribution from ${T_{\II,\II}}^{\mu\nu}$, 
it is not necessary to use the formulae 
(\ref{horizon1}) and (\ref{horizon2}), because this contribution 
is independent of the distance $\rho$ from the charge, and 
therefore we do not need to consider quantities in the region 
$H^+$. From Eqs.~(\ref{IIandII}) and (\ref{Killing}), we obtain
\begin{eqnarray}
\frac{\rho^2}{\X\cdot k}
\frac{4\pi}{e^2}{T_{\II,\II}}^{\mu\nu}X_\mu X_\nu
&=&
(\X\cdot k)
\left[
(a\cdot k)^2-a\cdot a
\right],
\label{hII_II}
\end{eqnarray}
for the region $\sigma(\eta)$ with arbitrary $\eta$.

In Eqs.~(\ref{hI_I}) and (\ref{hI_II}), we have evaluated the 
integrands in the region $H^+$. However, we should also evaluate 
these quantities in the region $I^+$, defined by the limit 
$\eta\rightarrow\infty$ along the light ray parallel to the $z$ 
axis. In this case, the limit $\eta\rightarrow\infty$ is 
equivalent to the limit $\rho\rightarrow\infty$.   
Estimating the $\rho$ dependence of 
$\rho^2{T_{\I,\I}}^{\mu\nu}X_\mu X_\nu$ 
and $\rho^2{T_{\I,\II}}^{\mu\nu}X_\mu X_\nu$ in this direction, 
we find that these quantities disappear in $I^+$.   
On the other hand, the right-hand sides of Eqs.~(\ref{hI_I}) and 
(\ref{hI_II}) are equal to zero for $k^\mu$ directed forward $I^+$, 
and we thus confirm that these equations can be extended to
the region $I^+$. 

Combining Eqs.~(\ref{hI_I})--(\ref{hII_II}), 
we obtain the angular dependence of the radiated power, 
\begin{eqnarray}
\lim_{\eta\rightarrow\infty}E_{[X,\sigma(\eta),{\rm ret}]}
&=&-\lim_{\eta\rightarrow\infty}
\int_{\sigma(\eta)} d\Sigma_\mu X_\nu {T_{\rm ret}}^{\mu\nu}
\nonumber\\
&=&\frac{e^2}{4\pi}d\tau
\oint d\Omega_0(\X\cdot k)
\left[
(\alpha\cdot k)^2-\alpha\cdot\alpha
\right],
\label{angular}
\end{eqnarray}
where 
$\alpha^\mu=h^\mu_{\ \nu}(a^\nu-w^\nu)=a^\mu-{w_\perp}^\mu$. 
By using the formulae
\begin{eqnarray}
\hspace{-1.5cm}\oint d\Omega_0 {k_\perp}^\mu=0, 
\ \ \ \ 
\oint d\Omega_0 {k_\perp}^\mu {k_\perp}^\nu {k_\perp}^\lambda =0, 
\ \ \ \ 
\oint d\Omega_0 {k_\perp}^\mu {k_\perp}^\nu=
\frac{4\pi}{3}h^{\mu\nu},
\end{eqnarray}
where ${k_\perp}^\mu=h^\mu_{\ \nu} k^\nu=k^\mu-v^\mu$,
the total radiated energy in the Rindler frame is obtained as
\begin{eqnarray}
\lim_{\eta\rightarrow\infty}E_{[X,\sigma(\eta),{\rm ret}]}&=&
\frac{2e^2}{3}\alpha^\mu\alpha_\mu(-\X\cdot v)d\tau.
\label{result}
\end{eqnarray}

At the instant that the charge is at rest in the Rindler frame 
(where $v^\mu=u^\mu$), we have $\alpha^\mu=a^\mu-g^\mu$. 
Therefore $\alpha^\mu$ should be interpreted as the acceleration 
of the charge relative to the Rindler frame.\cite{Hi01} 
We find from Eq.~(\ref{result}) that radiation is generated 
in the Rindler frame if and only if the charge deviates from the 
trajectory~\footnote{We here note the result obtained in 
Ref.~\citen{Hi01} that a charge governed by the equation 
$\alpha^\mu=0$ exhibits hyperbolic motion in an inertial frame, 
while it comes to rest in the Rindler frame in the infinite 
future.} $\alpha^\mu= 0$.

\section{Bound and emitted parts of the Maxwell tensor}
\label{BandE}

We have treated the situation in which the electromagnetic field 
surrounding the moving charge is expressed by the retarded field,
which is given as $[$see Eq.~(2.3) in Ref.~\citen{Te70}$]$
\begin{eqnarray}
{F_{\rm ret}}^{\mu\nu}&=&{F_{\ \I}}^{\mu\nu}+{F_{\ \II}}^{\mu\nu},
\nonumber\\
{F_{\ \I}}^{\mu\nu}&=&
\frac{e}{\rho^2}(v^\mu k^\nu-k^\mu v^\nu),\nonumber\\
{F_{\ \II}}^{\mu\nu}&=&
\frac{e}{\rho}a\cdot k(v^\mu k^\nu-k^\mu v^\nu)
+\frac{e}{\rho}(a^\mu k^\nu-k^\mu a^\nu),
\label{fsplit_iner}
\end{eqnarray}
where ${F_{\ \I}}^{\mu\nu}$ and ${F_{\ \II}}^{\mu\nu}$
behave as $\sim \rho^{-2}$ and $\sim \rho^{-1}$, respectively. 
In conventional treatments in inertial frames, the former part is 
interpreted as being bound to the charge, and the latter as being 
emitted from the charge. By using this splitting, Teitelboim 
split the energy momentum tensor into the form 
(see Eqs.~(\ref{IandI})--(\ref{IIandII}))
\begin{eqnarray}
{T_{\rm ret}}^{\mu\nu}&=&
{T_{\ \I}}^{\mu\nu}+{T_{\ \II}}^{\mu\nu},
\nonumber\\
{T_{\ \I}}^{\mu\nu}&=&
{T_{\ \I,\I}}^{\mu\nu}+{T_{\ \I,\II}}^{\mu\nu},
\ \ \ 
{T_{\ \II}}^{\mu\nu}={T_{\ \II,\II}}^{\mu\nu},
\label{Tei_splitting}
\end{eqnarray}
where the emitted part 
${T_{\ \II}}^{\mu\nu}$ is composed of the field 
${F_{\ \II}}^{\mu\nu}$, while the bound part 
${T_{\ \I}}^{\mu\nu}$ includes 
the pure ${F_{\ \I}}^{\mu\nu}$ part 
(denoted by ${T_{\ \I,\I}}^{\mu\nu}$) and the interference 
between ${F_{\ \I}}^{\mu\nu}$ and ${F_{\ \II}}^{\mu\nu}$ 
(denoted by ${T_{\ \I,\II}}^{\mu\nu}$).\cite{Te70}  

We now note that the radiation formula given in Eq.~(\ref{result}) 
resembles the Larmor formula, where the $a^\mu$ dependence in the 
latter is replaced by $\alpha^\mu$ dependence in the former. 
In analogy to the $a^\mu$ dependence found in the splitting 
(\ref{fsplit_iner}), we introduced the following splitting of the field 
for the Rindler frame 
$[$see Eq.~(2$\cdot$36) in Ref.~\citen{Hi01}$]$:
\begin{eqnarray}
{F_{\rm ret}}^{\mu\nu}&=&{F_{\ \tI}}^{\mu\nu}+{F_{\ \tII}}^{\mu\nu},
\nonumber\\
{F_{\ \tI}}^{\mu\nu}&=&
\frac{e}{\rho^2}(v^\mu k^\nu-k^\mu v^\nu)
+\frac{e}{\rho}\we\cdot k(v^\mu k^\nu-k^\mu v^\nu)
+\frac{e}{\rho}(\we^\mu k^\nu-k^\mu \we^\nu),
\nonumber\\
{F_{\ \tII}}^{\mu\nu}&=&
\frac{e}{\rho}\alpha\cdot k(v^\mu k^\nu-k^\mu v^\nu)
+\frac{e}{\rho}(\alpha^\mu k^\nu-k^\mu \alpha^\nu).
\label{fsplit_acc}
\end{eqnarray}
Here, the first part is independent of $\alpha^\mu$, while the second 
part is linear in $\alpha^\mu$. 
This splitting inplies the following splitting of the 
energy-momentum tensor into the emitted part 
${T_{\ \tII}}^{\mu\nu}$ and the bound part 
${T_{\ \tI}}^{\mu\nu}$ $[$Eq.~(2$\cdot$37) in 
Ref.~\citen{Hi01}$]$:
\begin{eqnarray}
{T_{\rm ret}}^{\mu\nu}&=&
{T_{\ \tI}}^{\mu\nu}+{T_{\ \tII}}^{\mu\nu},
\nonumber\\
{T_{\ \tI}}^{\mu\nu}&=&
{T_{\ \tI,\tI}}^{\mu\nu}+{T_{\ \tI,\tII}}^{\mu\nu}.
\end{eqnarray}
Here, the emitted part 
${T_{\ \tII}}^{\mu\nu}$ is composed of the field 
${F_{\ \tII}}^{\mu\nu}$, while the bound part 
${T_{\ \tI}}^{\mu\nu}$ includes 
the pure ${F_{\ \tI}}^{\mu\nu}$ part (denoted by 
${T_{\ \tI,\tI}}^{\mu\nu}$)
and the interference 
between ${F_{\ \tI}}^{\mu\nu}$ and 
${F_{\ \tII}}^{\mu\nu}$ (denoted by ${T_{\ \tI,\tII}}^{\mu\nu}$).
The explicit forms of these parts are 
\begin{eqnarray}
\hspace {-0.5cm}\frac{4\pi}{e^2}{T_{\ \tI, \tI}}^{\mu\nu}&=&
\frac{1}{\rho^6}
[1+(w\cdot r)^2+2(w\cdot r)-2\rho(w\cdot r)(v\cdot w)
-2\rho(v\cdot w)]r^\mu r^\nu\nonumber\\
\hspace {-0.5cm}& &
\hspace{-0.2cm}
-\frac{1}{\rho^5}(1+w\cdot r)[v^\mu r^\nu+r^\mu v^\nu]
-\frac{1}{\rho^4}[w^\mu r^\nu+r^\mu w^\nu]
-\frac{1}{2\rho^4}g^{\mu\nu},
\label{oneone}\\
\hspace {-0.5cm}
\frac{4\pi}{e^2}{T_{\ \tI, \tII}}^{\mu\nu}&=&
\frac{2}{\rho^6}[(\alpha\cdot r)\{1+w\cdot r-\rho(v\cdot w)\}
-\rho^2(w\cdot \alpha)]
r^\mu r^\nu\nonumber\\
\hspace {-0.5cm}
& &-\frac{1}{\rho^5}(\alpha\cdot r)[v^\mu r^\nu+r^\mu v^\nu]
-\frac{1}{\rho^4}[\alpha^\mu r^\nu+r^\mu \alpha^\nu],
\label{onetwo}\\
\hspace {-0.5cm}
\frac{4\pi}{e^2}{T_{\ \tII}}^{\mu\nu}&=&
\frac{1}{\rho^6}
[(\alpha\cdot r)^2-\rho^2(\alpha\cdot\alpha)]r^\mu r^\nu.
\label{twotwo}
\end{eqnarray}
Note that these expressions are independent of, linear in, and 
quadratic in $\alpha^\mu$, respectively. 
We here note that the emitted and bound parts of the 
energy-momentum tensor are conserved separately, 
\begin{eqnarray} 
\nabla_\mu{T_{\ \tI}}^{\mu\nu}=0, \ \ \ \ \  
\nabla_\mu{T_{\ \tII}}^{\mu\nu}=0,
\end{eqnarray} 
off the world line of the charge. $[$See Eqs.~(2$\cdot$38c) and 
(2$\cdot$39c) in Ref.~\citen{Hi01}. A simple derivation of these 
equations is given above  Eq.~(2$\cdot$26) 
in that reference.$]$ 

As mentioned in \S\ref{introo}, 
the validity of identificating these parts as the bound and emitted 
parts was confirmed by applying the three conditions introduced in 
Ref.~\citen{Hi01}. However, that was accomplished only after the 
angular integration of the Rindler energy over $\sigma(\eta)$. 
In the following, we show that the three conditions hold also 
for each element of solid angle. 

Condition 1 for each direction is confirmed trivially, because  
the tensor in Eq.~(\ref{twotwo}) itself disappears when
$\alpha^\mu=0$, which includes the case that the 
charge is fixed in the Rindler frame.\footnote{If one prefers 
to identify the radiation simply by the asymptotic definition 
without using the above splitting, it can be verified by 
noting that the integrand of Eq.~(\ref{angular}) disappears 
when $\alpha^\mu=0$.} 
We now consider Condition 2, using analysis similar to that used 
in the inertial case (see section 4 of Ref.~\citen{TV80}). 
For a 4-dimensional region $S$ with a boundary $\partial S$, 
we have a conservation law for the tensor satisfying
$\nabla_\mu T^{\mu\nu}=0$ within the region $S$,
\begin{eqnarray}
\int_{\partial S}d\Sigma_\mu X_\nu T^{\mu\nu}=0,
\label{dS}
\end{eqnarray} 
where $d\Sigma_\mu$ is the volume element of $\partial S$. 
This relation can be demonstrated using the Killing equation and 
Gauss's theorem  $[$see Eq. (2$\cdot$1) in Ref.~\citen{Hi01}$]$. 
Let us consider the segment $d\Omega_0(\eta_1)$ of 
$\sigma(\eta_1)$ and the segment $d\Omega_0(\eta_2)$ of 
$\sigma(\eta_2)$ constituting the same solid angle $d\Omega_0$. 
Then, applying the conservation law (\ref{dS}) to the 
region between $d\Omega_0(\eta_1)$ and $d\Omega_0(\eta_2)$ 
specified by different times $\eta_1\ne\eta_2$, and noting 
$X_\nu {T_{\ \tII}}^{\mu\nu}\propto k^\mu$, we obtain
\begin{eqnarray}
E_{[X,d\Omega_0(\eta_1),\tII]}
&=&
E_{[X,d\Omega_0(\eta_2),\tII]}.
\label{perangle}
\end{eqnarray}
This equation shows that Condition 2 holds for each element 
of solid angle. We can also confirm that the radiative energy 
in each direction is nonnegative, because 
$\rho^2(\alpha\cdot\alpha)-(\alpha\cdot r)^2\geq 0$ in 
Eq.~(\ref{twotwo}), and therefore the Rindler energy density of 
the part $\tII$ is always nonnegative.

Next, we examine Condition 3 for each direction. 
Except in the direction of the $z$ axis, this is simplified by 
using Eq.~(\ref{simple}) again. The angular dependence of the 
Rindler energy is given in Eq.~(\ref{sigmaeta}). Hence, by 
noting $X^\mu|_{H^+}\propto w^\mu$, we find that it is 
sufficient for our purpose to confirm 
${T_{\tI}}^{\mu\nu}|_{H^+}w_\mu w_\nu=0$. We can confirm this 
relation by contracting Eqs.~(\ref{oneone}) and (\ref{onetwo}) 
with $w^\mu$ and applying Eq.~(\ref{simple}). 
For the direction of the $z$ axis, we find 
$w^\mu r_\mu=0$, because of $w^\mu\propto r^\mu$. We can make 
the calculation somewhat simpler by using this relation and 
Eq.~(\ref{Killing}). Finally, we obtain 
$\rho^2{T_{\tI,\tI}}^{\mu\nu}X_\mu X_\nu\sim \rho^{-1}$
and $\rho^2{T_{\tI,\tII}}^{\mu\nu}X_\mu X_\nu\sim \rho^{-1}$ 
in the $z$ direction. Therefore, we find that these contributions 
disappear in the region $I^+$. From these results in 
$H^+$ and $I^+$, we find that Condition 3 holds for each 
element of solid angle.

\section{Conclusion}

We have obtained a simpler derivation of the radiation formula 
in the Rindler frame by using retarded coordinates, which are 
often used to simplify the integration of retarded 
quantities around a charge. Simplification also results from  
Eq.~(\ref{simple}) or Eq.~(\ref{horizon1}), with which we can 
characterize the distance between the charge and the future
horizon by the vector $w^\mu$, the vector which played a key role 
in the analysis of Ref~\citen{Hi01}.
In the retarded coordinates, we can also express the angular 
dependence of the radiation. Then, we have 
confirmed that the splitting of the energy-momentum tensor 
into the bound part $\tI$ and emitted part $\tII$ satisfies 
the three conditions (see \S{\ref{introo}}) also in each 
element of solid angle.   

\section*{Acknowledgements}
I wish to thank S.-Y.~Lin for helpful comments. I would also 
like to thank T.~Hara for helpful discussion.

\appendix

\section{\label{v_element}Retarded Coordinates}

In this appendix, we calculate the volume element of the 
3-dimensional plane $\Sigma$ orthogonal to the vector $N^\mu$ 
given in Eq~(\ref{volume}). We start with describing the general 
properties of the retarded coordinates.\cite{TV80,Ta76}

The direction of $k^\mu$ is invariant when $\tau$ is varied with 
$\theta$ and $\varphi$ fixed. This invariance is expressed as
$\partial k^\mu/\partial \tau\propto k^\mu$. This property and 
differentiation of $k\cdot v=-1$ with respect to $\tau$ lead to
\begin{eqnarray}
\frac{\partial k^\mu}{\partial\tau}&=&(k\cdot a)k^\mu,
\label{diff_k}
\end{eqnarray}  
where $a^\mu=dv^\mu/d\tau$ is the 4-acceleration of the charge.
Using this, the total derivative of
$x^\mu=\bx^\mu+\rho k^\mu$ gives
\begin{eqnarray}
dx^\mu &=& [v^\mu+\rho (r\cdot a)k^\mu]d\tau+k^\mu d\rho
+\rho[k_\theta^\mu d\theta+k_\varphi^\mu d\varphi],
\label{dx}
\end{eqnarray}
where $k^\mu_\theta=\partial k^\mu/\partial\theta$ and
$k^\mu_\varphi=\partial k^\mu/\partial\varphi$.

From $k\cdot k=0$ and $k\cdot v=-1$, 
we obtain the orthogonality relations
$k_\theta\cdot k= k_\varphi\cdot k= 
k_\theta\cdot v= k_\varphi\cdot v=0$.
We can also set $k_\theta\cdot k_\varphi=0$, since from 
Eq.~(\ref{diff_k}) we get
\begin{eqnarray}
\frac{\partial}{\partial\tau}(k_\theta\cdot k_\varphi)
&=&2(k\cdot a)(k_\theta\cdot k_\varphi),
\end{eqnarray}
so that, if we set $(k_\theta\cdot k_\varphi)=0$ at one time, 
it holds for all time. Now, we also set 
$k^\mu_\perp=h^\mu_{\ \nu} k^\nu=k^\mu-v^\mu$, where
$h^\mu_{\ \nu}=\delta^\mu_{\ \nu}+v^\mu v_\nu$ is the projector 
onto the plane orthogonal to $v^\mu$. We find that 
$k_\perp\cdot k_\perp=1$ and that $v^\mu$, 
$k^\mu_\perp$, $k^\mu_\theta$ and $k^\mu_\varphi$ constitute an 
orthogonal basis.

We now calculate the volume element of the 3-dimensional 
plane $\Sigma$. For any displacement within the plane, we find 
$N^\mu dx_\mu=0$. From this and Eq.~(\ref{dx}), we obtain
\begin{eqnarray}
d_\tau x^\mu&=&
\left(
v^\mu-\frac{N\cdot v}{N\cdot k}k^\mu
\right)d\tau,\nonumber\\
d_\theta x^\mu&=&
\left(
k^\mu_\theta-\frac{N\cdot k_\theta}{N\cdot k}k^\mu
\right)\rho d\theta,\nonumber\\ 
d_\varphi x^\mu&=&
\left(
k^\mu_\varphi-\frac{N\cdot k_\varphi}{N\cdot k}k^\mu
\right)\rho d\varphi,
\label{displacements}
\end{eqnarray}
where $d_\tau x^\mu$ is the displacement within $\Sigma$ when 
$\tau$ alone varies, and $d_\theta x^\mu$ and $d_\varphi x^\mu$ 
are displacements within $\Sigma$ when $\theta$ and $\varphi$ 
alone vary, respectively. The volume element $d\Sigma_\mu$ of 
$\Sigma$ is given by
\begin{eqnarray}
d\Sigma_\mu&=&
\epsilon_{\mu\nu\lambda\rho}
d_\tau x^\nu d_\theta x^\lambda d_\varphi x^\rho, 
\label{define}
\end{eqnarray}
where $\epsilon_{\mu\nu\lambda\rho}$ is the Levi-Civita 
permutation symbol. From Eq.~(\ref{displacements}), we have
\begin{eqnarray}
d\Sigma_\mu&=&
\rho^2 d\tau d\theta d\varphi
\left(
\epsilon_{\mu\nu\lambda\rho}v^\nu k^\lambda_\theta k^\rho_\varphi
-\frac{N\cdot k_\varphi}{N\cdot k}
\epsilon_{\mu\nu\lambda\rho}v^\nu k^\lambda_\theta k^\rho
\right.\nonumber\\
& &\hspace{2.3cm}\left.
-\frac{N\cdot k_\theta}{N\cdot k}
\epsilon_{\mu\nu\lambda\rho}v^\nu k^\lambda k^\rho_\varphi
-\frac{N\cdot v}{N\cdot k}
\epsilon_{\mu\nu\lambda\rho}k^\nu k^\lambda_\theta k^\rho_\varphi
\right).
\end{eqnarray}
The first term on the right-hand side of this relation is orthogonal 
to $v^\mu$, $k_\theta^\mu$ and $k_\varphi^\mu$, 
and therefore it is proportional to $k_\perp^\mu$. 
Similarly, the second, third and fourth terms are proprtional 
to $k_\varphi^\mu$, $k_\theta^\mu$ and $k^\mu$ respectively. 
Proportionality factors are obtained by contraction with 
$k_\perp^\mu$, $k_\varphi^\mu$, $k_\theta^\mu$ and $k^\mu$. 
We find 
\begin{eqnarray}
\hspace {-1.2cm}
d\Sigma^\mu&=&
\rho^2 d\tau d\Omega_0
\left[
-\frac{k_\perp^\mu}{k_\perp\cdot k_\perp}
-\frac{N\cdot k_\varphi}{N\cdot k}
\frac{k_\varphi^\mu}{k_\varphi\cdot k_\varphi}
-\frac{N\cdot k_\theta}{N\cdot k}
\frac{k_\theta^\mu}{k_\theta\cdot k_\theta}
+\frac{N\cdot v}{N\cdot k}k^\mu
\right],
\label{expanded}
\end{eqnarray}
where
\begin{eqnarray}
d\Omega_0&=&
\epsilon_{\lambda\mu\nu\rho}
v^\lambda k^\mu k_\theta^\nu k_\varphi^\rho d\theta d\varphi
\label{solid}
\end{eqnarray}
is the solid angle element for the inertial frame with time axis 
$v^\mu$. In Eq.~(\ref{define}), the volume element $d\Sigma^\mu$ is 
assumed to be orthogonal to the plane $\Sigma$. This can be verified 
by rewriting $N^\mu$ in terms of the orthogonal basis $v^\mu$, 
$k^\mu_\perp$, $k^\mu_\theta$ and $k^\mu_\varphi$, and comparing 
the result with Eq.~(\ref{expanded}). Finally, we obtain 
Eq.~(\ref{volume}).


\begin{thebibliography}{99}
\bibitem{Hi01}
T.~Hirayama, 
Prog.\ Theor.\ Phys. {\bf 106} (2001), 71, gr-qc/0102082.
\bibitem{Ri66}
W.~Rindler, Am.\ J.\ Phys.\ {\bf 34} (1966), 1174.
\bibitem{PV99}
This point of view was discussed in analogy with the quantum 
theory (Unruh effect)\cite{BD82} in M.~Pauri and M.~Vallisneri, 
Found.\ Phys.\ {\bf 29} (1999), 1499, gr-qc/9903052.
\bibitem{BD82}
N.~D.~Birrell and P.~C.~W.~Davies, {\it Quantum Fields in 
Curved Space} 
(Cambridge University Press, Cambridge, 1982).
\bibitem{FR60}
T.~Fulton and F.~Rohrlich, 
Ann.\ of\ Phys.\  {\bf 9} (1960), 499.
\bibitem{Bo80}
D.~G.~Boulware, Ann.\ of\ Phys.\  {\bf 124} (1980), 169.
\bibitem{Ro61}
F.~Rohrlich, Nuovo\ Cim.\ {\bf 21} (1961), 811.
\bibitem{Te70}
C.~Teitelboim, Phys.\ Rev.\ D~{\bf 1} (1970), 1572.
\bibitem{TV80}
C.~Teitelboim, D.~Villarroel and Ch.~G.~Van~Weert, 
Riv.\ Nuovo\ Cim. {\bf 3}, No. 9 (1980). 
\bibitem{Ta76}
R.~Tabensky, Phys.\ Rev.\ D~{\bf 13} (1976), 267.
\end{thebibliography}
\end{document}